\documentclass[aps,prl,reprint,amsmath,amssymb,twocolumn]{revtex4-2}

\usepackage{graphicx}
\usepackage{dcolumn}
\usepackage{bm}
\usepackage{xcolor}
\usepackage{amsmath}
\usepackage{psfrag}
\usepackage[normalem]{ulem}

\newcommand{\beq}{\begin{equation}}
\newcommand{\eeq}{\end{equation}}

\newcommand{\brac}[1]{\left({#1}\right)}
\newcommand{\pd}[2]{\frac{\partial{#1}}{\partial{#2}}}

\newcommand{\curl}{\nabla\times}
\renewcommand{\div}{\nabla\cdot}

\newcommand{\bB}{{\boldsymbol B}}

\newcommand{\bv}{{\boldsymbol v}}

\def\apj{Astrophys. J.}
\def\apjl{Astrophys. J. Lett.}

\def\aap{Astron. Astrophys. }

\def\mnras{Mon. Not. Roy. Astron. Soc. }

\def\nat{Nature}
\def\prl{Phys. Rev. Lett.}
\def\prd{Phys. Rev. D.}

\def\prc{Phys. Rev. C.}

\newcommand{\skl}[1]{{\color{black}{#1}}}

\begin{document}

\title{\skl{Signatures of magnetic flux expulsion from neutron star cores}}

\author{S. K. Lander${}^1$}
\email{samuel.lander@uea.ac.uk}
\author{K. N. Gourgouliatos${}^2$}
\author{Z. Wadiasingh${}^{3,4,5}$}
\author{D. Antonopoulou${}^6$}
\affiliation{${}^1$School of Engineering, Mathematics and Physics, University of East Anglia, Norwich, NR4 7TJ, U.K.,}
\affiliation{${}^2$Laboratory of Universe Sciences, Department of Physics, University of Patras, Patras, Rio 26504, Greece}
\affiliation{${}^3$Department of Astronomy, University of Maryland, College Park, Maryland 20742, USA, }
\affiliation{${}^4$Astrophysics Science Division, NASA Goddard Space Flight Center,Greenbelt, MD 20771, USA, }
\affiliation{${}^5$Center for Research and Exploration in Space Science and Technology, NASA/GSFC, Greenbelt, Maryland 20771, USA,}
\affiliation{${}^6$Jodrell Bank Centre for Astrophysics, Department of Physics and Astronomy, The University of Manchester, UK}

\begin{abstract}
Shortly after a neutron star is born, the protons in its core
  begin to form a superconductor. In terrestrial materials, the
  hallmark of superconductivity is an associated expulsion of magnetic
  flux, but whether this expulsion process can be effective in neutron
  stars remains an open question -- one with major implications
  for the phenomenology of pulsars and magnetars.
Earlier theoretical arguments suggested flux must be trapped within the core,
yet models of magnetars rely on it being expelled from the core and
confined to the crust, where it can evolve on kyr timescales. We show
that if expulsion is not complete, a qualitatively new evolutionary
branch for neutron stars arises, which can account for the properties
of newly discovered long-period radio transients and fast radio bursts
in older environments. One recently proposed model that could create such field topologies has additional implications for gravitational wave emission and predicts a characteristic energy release that, if observed, will corroborate the role of reconnection at the onset of superconductivity and can constrain the superconducting proton gap. 

\end{abstract}

\maketitle

\label{firstpage}


\maketitle


Neutron star (NS) cores host the only known superconductor in the
Universe outside terrestrial laboratories \cite{hask_sed}. Superconductivity does not set in simultaneously across the whole NS core because of the strong radial dependence of the critical onset temperature $T_c$. The transition begins in a thin shell in the star's outer core which slowly expands both outwards (stopping at the boundary with the crust) and inwards, as more regions cool below $T_c$. 
If the local magnetic field $\bB$ is weaker than a critical field $H_c$, the minimum-energy state in the superconducting shell will be dictated
by the \emph{Meissner effect}: the expulsion of magnetic flux, leaving
$B=0$ except in a surface layer \cite{tinkham}.
\begin{figure}
\begin{center}
\begin{minipage}[c]{0.9\linewidth}
\includegraphics[width=\linewidth]{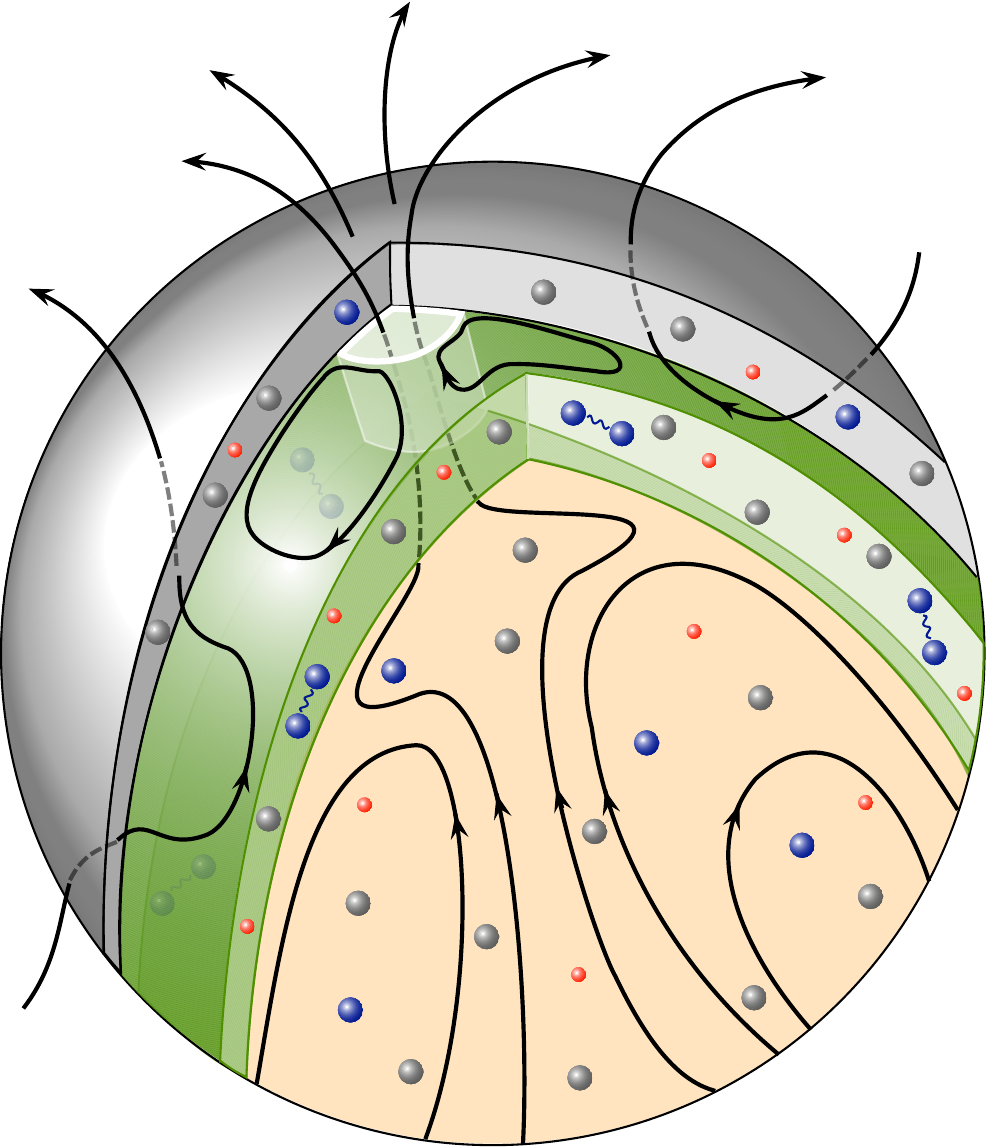}
\end{minipage}
\caption{\label{sc_onset}
A neutron star is mainly composed of neutrons, protons and electrons
(grey, indigo and red balls, respectively). The magnetic field
(black lines with arrows) is sourced by a relative flow of protons and electrons. Minutes after the star's
birth, protons in a layer (green) in the outer core become
superconducting. Unless $B>H_c$, the star's minimum-energy state requires the field to be expelled from
this layer: the Meissner effect. This expulsion can however be 
incomplete -- e.g. here we sketch a polar hole where $B\neq 0$. Such
field configurations have several interesting observational implications, as we will discuss.}
\end{center}
\end{figure}
Whilst on Earth this process is well understood and near-instantaneous,
NSs present a different environment. Their intense $B$ (typically
$10^{12}-10^{15}\,\mathrm{G}$), is intrinsic to the star, the
  result of a persistent relative flow between the protons and
  electrons
(in a background of free neutrons), inside which the superconductor forms and
expands; see Fig. \ref{sc_onset}. In the laboratory case only electrons can become superconducting; in NSs it is the proton fluid. The strong gradients in density $\rho$ and in $T_c$, together
with the temperature dependence of $H_c$, lead to a broad
normal-superconducting transition; there is no static equilibrium for
the system to reach for at least decades, and potentially far longer
\cite{ho17}.

A long-standing uncertainty about whether effective flux expulsion
is possible in NSs has led to a major tension in the literature. 
Microscopic arguments \cite{BPP69a} suggest that the process is so
slow that the flux will be trapped inside the core and quantised into
thin fluxtubes, even if $B<H_c$. Models of pulsar rotational evolution
and spin-up glitches often explicitly or implicitly assume that the
magnetic field threads the entire star \cite{hask_mel,ant_HE}. On the
other hand, models that explain magnetar activity with
Hall-drift-driven crustal field evolution
(e.g. \cite{hollrued02,ponsgepp07}), require a $B=0$ inner
(crust-core) boundary condition, i.e. that flux is completely expelled
from the core, without giving any detailed justification for this
assumption.  
 The uncertainty stems from the fact that the Meissner `effect' is simply a statement about the minimum-energy
equilibrium state for the system, and does not dictate the process by which this is reached.
Recent research examining the gradual transition to superconductivity
in the NS core proposed a new mechanism that could realise flux expulsion on astrophysically relevant timescales \cite{meissner_long}. In addition to complete flux expulsion from the outer core, which can potentially resolve the tension between previous theoretical work and magnetar
modelling, another possible outcome is a $B=0$ shell that is broken by one
or two large holes penetrated by $\bB$ (see below and
\cite{meissner_long} for details). In this article we examine some important observational
consequences of such a model: a
late-time energy injection following a supernova, changes to the mature star's rotational properties, and a revised prognosis for the detection of continuous gravitational waves from pulsars like the Crab. We moreover compare simulations of crustal magnetic field
evolution for full Meissner expulsion with two different partially-expelled
geometries. Whilst the latter are inferred from the model in \cite{meissner_long}, such states
are conceivable endpoints of any operative Meissner effect in NSs, and
present a qualitatively new evolutionary path: late-onset and
  protracted magnetic activity in the crust. This delayed activation of old NSs could solve the mystery of some recently-discovered long-period radio transients \cite{caleb_discovery_2022,hurley-walker_galactic_2022,2024arXiv240707480D,2024arXiv241116606W,2025NatAs.tmp...36L} and FRBs associated with aged host environments \citep{2022Natur.602..585K,2024ApJ...967...29L,2025ApJ...979L..21S,2025ApJ...979L..22E}.   

\section{The onset of superconductivity}

The production of a macroscopic $B=0$ region in the core has typically been
associated with very slow
mechanisms for flux transport (e.g. \cite{BPP69a,jones91,koch96,graber15}), i.e. implicitly assuming that the process to achieve Meissner expulsion must be secular and
dissipative. Expulsion via Ohmic decay, for example, will take
  $\sim 10^9\,\mathrm{yr}$ for a field whose characteristic
  lengthscale is the stellar radius $R_*\approx 10\,\mathrm{km}$ \cite{BPP69a,BPP69b}.
Such slow expulsion suggests that magnetic fluxtubes will thread the
interior of every observed NS -- even in the likely case where
$B<H_c$. A comparison of cooling and (Ohmic) flux dissipation
timescales concluded that for NSs younger than $10^6$~yr, any
field-free regions will likely be $\lesssim 10^3\,\mathrm{cm}$ in
extent \cite{ho17}. However, the Meissner effect is not intrinsically
dissipative, leaving open other possibilities for flux expulsion.

\skl{A universal obstacle for any flux-expulsion mechanism is that}
the nascent shell of superconductivity will be crossed by field lines,
\skl{and so creating a $B=0$ shell without substantial flux
  rearrangement would produce radial discontinuities in the field,
  which }violate $\div\bB=0$. The radial field could, however, be
rearranged into an angular component through advection via a suitable
fluid flow $\bv$:
\beq
\pd{\bB}{t}=\curl(\bv\times\bB),
\eeq
where $|\bv|$ is only limited by the (high) local Alfv\'en speed.
If this results in a suitably pinched field geometry, reconnection can
then act to produce a $B=0$ shell, provided that these processes are faster than the
spreading of the superconducting region.
%
%
Advection of this form was recently explored for a simplified
axisymmetric model \cite{meissner_long}, which argued -- by comparing cooling and reconnection timescales, and using the restriction that for $B>H_c$
flux expulsion is not energetically favourable -- that dynamical Meissner expulsion is \emph{possible} only when
\beq
10^{12}\,\mathrm{G}\lesssim B\lesssim 5\times 10^{14}\,\mathrm{G}.
\eeq
Even within this range, the field geometry and fluid flow need not be conducive to
complete reconnection. \skl{A complex, small-scale field geometry is
  the most likely to undergo effective reconnection, perhaps requiring
  just vestigial fluid motions remaining from the supernova
  \cite{meissner_long}.
  By contrast, the most challenging geometry to expel
  is probably an axisymmetric dipolar field, because this would require a
  km-scale fluid circulation whose radial component would be
  resisted by the star's stable stratification. Nonetheless, it is a
  geometrically simple first approximation that allows us
  to set limits on when an
  advection-driven Meissner effect may occur. As well as the possible
  outcomes of full or no flux expulsion,} shearing and
reconnection \skl{of $\bB$} at the incipient shell
\skl{of superconductivity} can lead to a broken
$B=0$ superconducting shell pierced with field lines through either
two roughly cylindrical $B\neq 0$ holes (one at each pole), or one
equatorial band \cite{meissner_long}. The size of these $B\neq 0$
regions in the shell may be calculated from conservation of 
magnetic flux
and the requirement that
$B\leq H_c$; the result is holes whose cross-sectional area is
proportional to $B$.
Whether or not this $B=0$ shell is complete, however, there must
always be an inner trapped core of magnetic field that will be
compressed until it reaches $B=H_c$. Magnetic
flux conservation then yields an estimate of the final radius $R_{\mathrm{in}}$ of this inner core:
\beq\label{Rcore}
R_{\mathrm{in}}\approx R_*\sqrt{B/H_c}.
\eeq
\skl{Within the advective model of \cite{meissner_long} a full range
  of more complex
  flux distributions are also possible, depending on the
  pre-superconducting field geometry and the fluid flows. As there may
  also be different mechanisms for producing anisotropic flux distributions,
  in what follows we examine a few generic outcomes, and discuss ways to test and constrain these
  observationally.}

\section{Results}

We first examine \skl{one} specific prediction of the advection-driven
expulsion mechanism \cite{meissner_long}, which might render the model
observationally testable. \skl{We then look at observational
  manifestations of NSs with and without flux expulsion. Independently
  of the expulsion mechanism, there is an} intriguing possibility of macroscopic $B=0$ regions and possibly of
more complex geometry than a fully flux-free NS core. Such field
topologies will have major ramifications for both electromagnetic and
gravitational radiation from NSs, but also for their magnetorotational
evolution. We numerically investigate the long-term magnetic field
evolution under both the typical `magnetar' $B=0$ crust-core boundary
condition, implying full Meissner expulsion in the outermost core
region, as well as \skl{two} partially-expelled geometries, \skl{as} discussed
above. The results are especially important in view of recent
discoveries that challenge NSs as the source of some high-energy
transients. Last, we touch upon the rotational implications of such
field topologies.

\subsection{Energy release at superconductivity onset}

\skl{In the particular case of dynamical expulsion, as proposed in
  \cite{meissner_long}, an electromagnetic signature of flux expulsion
is expected at the onset of superconductivity.} The kind of shearing motion needed to \skl{bring} $\bB$ into a
geometry amenable to reconnection causes an approximate
doubling of a typical field line's length and, therefore, of the
magnetic energy, \skl{in the axisymmetric
  dipole case}; see \cite{meissner_long} for detailed calculations. \skl{Such an increase in magnetic energy, in order to
  minimise the total \emph{free} energy of the system, is a known
  property of superconducting systems \cite{annett}.} But any subsequent
reconnection event would then return $\bB$ to a less-sheared, lower-energy 
configuration. \skl{Two very different recent examples of such
  potential flux-expelled end-state equilibria are the microscopic
  calculations of \cite{wood_grab} and the global calculations of
  \cite{das25}. In both cases, however, a dynamical mechanism is
  needed to realise the equilibrium state; if this happens at the
  onset of superconductivity, we expect} a sudden
magnetic energy release which, \skl{for an axisymmetric dipolar geometry}, may be as large as
$10^{46}-10^{47}$ erg, \skl{but for more complex and
  small-scale field geometries is likely weaker} \cite{meissner_long}.

This reconnection energy release in the core will cause local heating and shake field lines that extend through the still-molten crust and out to the magnetosphere. If this complex transfer of magnetic energy is
radiatively efficient in producing a detectable signal, the reconnection event could ultimately be seen
as a peculiar transient, possibly involving delayed energy injection
into the supernova remnant. Given the vast zoo of optical transients
\cite{2014ApJ...794...23D}, such signals may
already have been detected. Alternatively, the reconnection event might be
more readily identifiable if a binary NS merger produces a
massive magnetized NS remnant: a cleaner environment for
signals to transit. Many short gamma-ray bursts exhibit X-ray extended
emission (lasting $\sim10^2-10^3$~s), whose physical origin is
uncertain and requires delayed energy injection
\cite{2015MNRAS.452..824K}. Secure association of such signals would allows us to constrain the maximum
value of $T_c$ and therefore the proton pairing gap model.

\subsection{Continuous gravitational wave emission}
Recent studies on the non-detection of continuous gravitational waves
from various pulsars have put some stringent and
physically-interesting bounds on their ellipticity $\epsilon$
\cite{eps_crab_LIGO_08}. For example, the
Crab pulsar has $\epsilon\lesssim 3\times 10^{-5}$ \cite{eps_max_LIGO_22}, and this limit will continue to drop with advances in gravitational-wave interferometry. The Crab's inferred dipole surface magnetic field $B_{\rm dip}=4\times 10^{12}\,\mathrm{G}$. If its $\epsilon$ is primarily due to an average interior field
$\bar{B}_\mathrm{int}$ threading the entire core, ellipticity
calculations for superconducting NSs \cite{L14} allow us to infer that
\beq
\bar{B}_{\mathrm{int}}\lesssim 10^{15}\,\mathrm{G}\approx 300B_{\rm dip}.
\eeq
We can also invert this logic to assess the prospects
for a future gravitational-wave detection from the Crab, and how the
Meissner effect alters this prognosis. A typical NS could plausibly
\cite{igoshev21} harbour a field $\bar{B}_{\rm int}\gg
B_{\mathrm{dip}}$; let us take, for example,
$\bar{B}_{\mathrm{int}}=20B_{\mathrm{dip}}$ for the Crab, close to
typical magnetar values. If the Meissner effect has expelled the field
from a complete shell of the star, it will cease once an inner core of
$\bar{B}=H_c,R_{\rm in}\approx 0.4R_*$ (see equation \eqref{Rcore})
has been formed. However, the ellipticity for a superconducting core
scales as $\epsilon\propto \bar{B} R_{\mathrm{in}}^3$, so although
$\bar{B}$ has been increased substantially through flux compression,
the overall effect on $\epsilon$ is dominated by the decrease in
$R_{\mathrm{in}}$. As a result, the ellipticity for the
Meissner-expelled model $\epsilon_M$ is substantially lower than that
of the non-Meissner model $\epsilon_{nM}$ (i.e. where field lines
thread the entire core):
\begin{equation}
    \epsilon_M\approx 3\times 10^{-7},\epsilon_{nM}\approx 2\times 10^{-6}.
\end{equation}
In this scenario, if the Crab hosts a Meissner-expelled region, we
would expect weak gravitational-wave emission, but notable
magnetar-like activity (given that a crustal field of strength
$20B_{\mathrm{dip}}$ would have evolved substantially over the
lifetime of the Crab). The latter has, however, not been seen from the Crab
despite over fifty years of monitoring, leading us to conclude that
\emph{if} the Crab has a strong $\bar{B}_{\rm int}$ it must thread the entire core. Its associated gravitational-wave signal
  would then be relatively strong; below the sensitivity of current instruments by a
  factor of $\sim 10$, but potentially detectable by next-generation
instruments.

\subsection{Crustal field evolution}

\begin{figure*}
\begin{center}
\begin{minipage}[c]{0.7\linewidth}
\includegraphics[height=6cm]{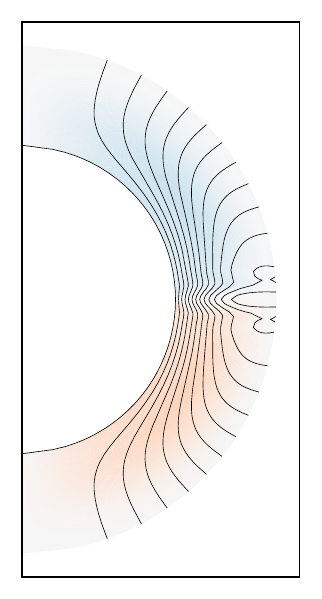}
\includegraphics[height=6cm]{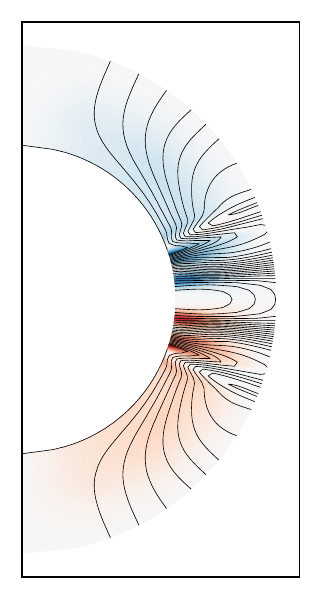}
\includegraphics[height=6cm]{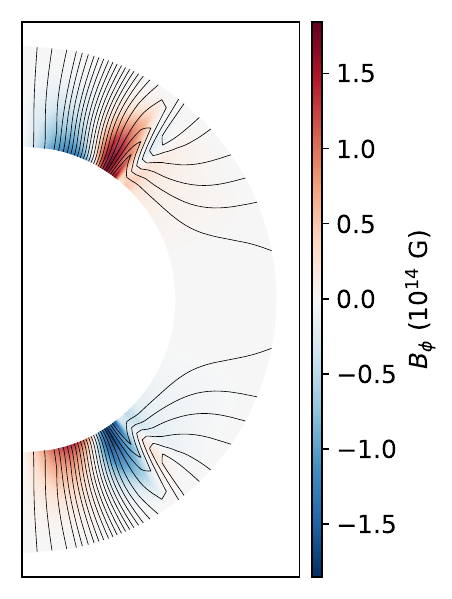}
\end{minipage}
\caption{\label{crust_B}
Magnetic field in the crust (here scaled up in thickness by a factor of 8 for
clarity) after 0.1 Myr of evolution, where the inner boundary at the base of the crust is: $B=0$ everywhere (left), field lines penetrating into the core in a band around the equator (middle), field lines penetrating into the core through two polar holes (right). Poloidal field lines are in black, toroidal field magnitude is shown with the colourscale.
}
\end{center}
\end{figure*}

In the absence of flux expulsion, $\bB$ is expected to
penetrate from the crust into the core \citep{BPP69a}, which should be
incorporated as a $B\neq 0$ inner boundary condition in numerical
evolutions of the crustal magnetic field. The result is a slowly
evolving crustal field without any particularly dramatic features
\cite{vigano13}. By contrast, the commonly employed $B=0$ inner 
boundary condition results, for young NSs
($10^3-10^4$ yr) with crustal fields $\gtrsim 10^{13}\,\mathrm{G}$, in
an active Hall drift that produces local patches of intense field;
conversely, for older sources the Hall effect saturates \cite{gour_cumm} and
eventually Ohmic decay dominates, smoothing out
small-scale features and weakening the field (see
e.g. \cite{ponsgepp07}). This fits with the theoretical picture of
canonical magnetars as young and highly magnetised NSs whose activity
is powered by crustal magnetic-field evolution; a $B=0$ inner boundary
is however critical for producing such activity. The dynamical Meissner
effect \cite{meissner_long} is the first scenario to provide a physically plausible
justification for this $B=0$ condition. However, as discussed above, it also allows
for other crust-core boundary conditions, where the
$B=0$ shell is broken by regions threaded by field lines: in the simplest cases, either an equatorial band or two polar holes. Figure \ref{crust_B} compares the result of 0.1 Myr of evolution of an initially poloidal field with surface strength $5\times 10^{13}\,\mathrm{G}$, for these three classes of boundary condition. With the usual $B=0$ inner
boundary (left-hand panel), the crustal field at this late stage is smooth, quasi-stationary, and relatively weak: the maximum value of the toroidal field $B_\phi^{\rm
  max}=4\times 10^{13}\,\mathrm{G}$.
By contrast, the simulation with $B=0$ around the poles but $B\neq 0$
in an equatorial band (middle panel) remains very dynamic, displaying sharp features in the poloidal field lines and
$B_\phi^{\rm max}=1.6\times 10^{14}\,\mathrm{G}$ close to the
transition region between the $B=0$ and $B\neq 0$ surfaces. Above this
region, small closed magnetic-field loops are
progressively pushed out of the star. All these features, which are
shared with the corresponding simulation featuring polar $B\neq 0$
holes (right-hand panel), are qualitatively new and suggest the
existence of a class of NS that becomes more active at $\sim 100$
times the age of conventional magnetars.
In the virtually untwisted magnetospheres of such old NSs,
  localised disruptions -- like those arising from our simulations --
  might manifest themselves through radio emission \citep{cooper_wad};
  see Fig. \ref{observables}.

In fact, such qualitatively different `old magnetars' may already have
been observed. The recent unexpected discoveries of very long-period intermittent
radio transients
\cite{hurley-walker_galactic_2022,2024arXiv241116606W,2025NatAs.tmp...36L}
 present a theoretical challenge: whilst a promising explanation is
 that of late-time activity of highly-magnetised NSs, \skl{especially
   given magnetar-like inferred field strengths for two of them \cite{caleb_discovery_2022,2024arXiv240707480D},} there is no clear mechanism to drive such behaviour within the standard models of NS magnetism \cite{beniamini_evidence_2023}.
Similarly, this new possible population of NSs could be behind the emission of some individual repeating FRBs that are associated with old sources/environments \citep{2021ApJ...908L..12T}, given their apparent non-trivial delay time with respect to star formation in their host galaxies \citep{2022Natur.602..585K,2024ApJ...967...29L,2025ApJ...979L..21S,2025ApJ...979L..22E}. There are several indications of a connection between the long-period radio transients and FRB-like emitting sources \citep{beniamini_periodicity_2020,doi:10.1126/sciadv.adp6351,men25}; the `old magnetar' evolutionary path, that opens if partial Meissner expulsion is possible, naturally links these phenomena to our understanding of NS populations. 
Finally, such NSs could also manifest in some high-mass X-ray binaries, as standard accretion models for the formation of long-period NSs also appeal to the existence of sustained magnetar-like fields for
$\gg 10^4\,\mathrm{yr}$ \cite{1999ApJ...513L..45L}.

\subsection{Rotational dynamics}

Most models of NS rotation (in contrast with magnetar modelling) assume a continuous $\bB$ threading the core and connecting it to the crust. This provides fast electromagnetic coupling between all charged core components and the crust, irrespective of whether the protons are superconducting, which justifies treating them as being in co-rotation. If the neutrons are normal, they will also be strongly coupled:
either near-instantaneously via the strong interaction (for normal protons), or due to particle collisions (for superconducting protons) -- still a relatively rapid process compared with most other timescales of interest, e.g. spin-down (see \cite{1984ApJ...282..533A,hask_sed} and references therein).

The presence of a macroscopic $B=0$ core region could drastically change this picture. 
Spin-down is then likely achieved by the
formation of a viscous crust-core boundary layer. The
associated Ekman flow should reduce the coupling timescale
compared to that of viscous diffusion, although realistic modelling
remains challenging. It is conceivable that a magnetically-decoupled
region (the $B=0$ shell and, for full Meissner expulsion, the inner
$B\neq 0$ core too) would deviate from `rigid’ co-rotation with the
crust (cf. \cite{2012ApJ...761...32M, 2015MNRAS.450.1638G} for core
superrotation in other setups); this problem deserves a careful
treatment (including, e.g., the various interfaces) to assess whether
it could lead to a long-lived rotational lag, or affect the response
to fast spin changes, like pulsar glitches. 

Once the neutron fluid becomes superfluid, it predominantly couples to the charged component via its rotational quantised-circulation vortices. These vortices become strongly magnetised due to entrainment if protons are superconducting, leading to electron-vortex relaxation timescales of $\sim 1-10$ sec \cite{1984ApJ...282..533A}. Thus the core superfluid maintains
differential rotation but is close to corotation and spins down with
the local charged component.
If the flux is not expelled but instead is confined in fluxtubes, then their interaction with vortices must be taken into account. This has been extensively studied in the literature
(e.g. \cite{2020MNRAS.493..382S}), mostly in relation to mode damping,
arguments against long-period NS precession
\cite{2003PhRvL..91j1101L}, and the angular momentum reservoir
available to generate large glitches as in the Vela pulsar
\cite{2012PhRvL.109x1103A}. These results
could change drastically in a $B=0$ region, since
fluxtubes will no longer co-exist with
neutron vortices there, and so
should be re-evaluated for
the different Meissner scenarios discussed
here. For example, in the case of a $B\neq 0$ equatorial band vortices
could naturally pin to a different number of fluxtubes depending on
the pulsar, giving rise to different glitch signatures
\cite{2020MNRAS.493L..98S}. By contrast, in the fully
Meissner-expelled or polar-hole case, vortices might not be pinned at
all, resolving contradictions with long-period precession and some
glitch models \cite{2006A&A...458..881L,2013ApJ...764L..25H}.
Finally, because vortices move outwards as a NS spins down, it
  has long been thought \cite{1998ApJ...492..267R} that their
  pinning to fluxtubes could transport magnetic flux outwards
  too. The precise nature of this secular `expulsion' in old NSs will
  be directly affected by the different Meissner scenarios in the
  star's early life.

\section{Discussion}

\begin{figure*}
\begin{center}
\begin{minipage}[c]{0.8\linewidth}
\includegraphics[width=\linewidth]{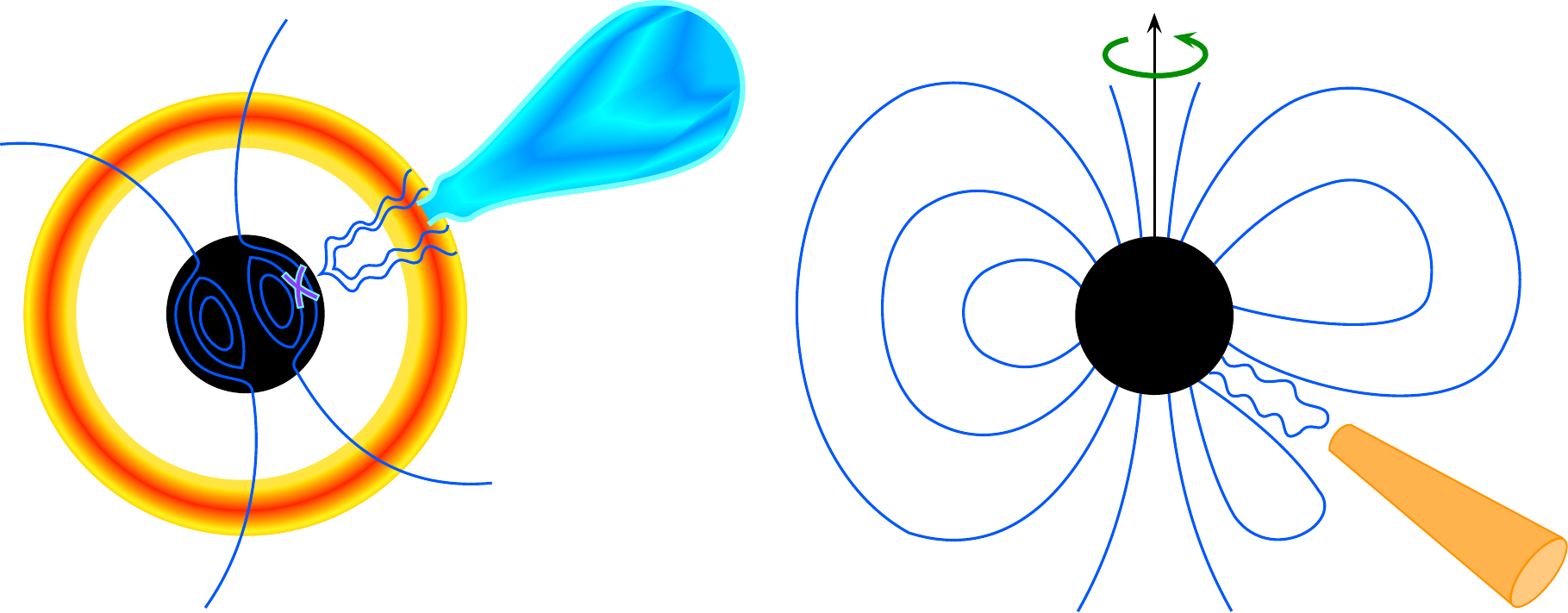}
\end{minipage}
\caption{\label{observables}
Two possible ways to `observe' the Meissner effect in a neutron star
(black circle) via disruptions to its external magnetic-field
lines. Left: at an age of a few minutes, the reconnection of
magnetic field lines (blue) -- e.g. as may be required to create a magnetar
-- releases a huge amount of
energy and shakes field lines crossing the nebula (red-yellow disc),
potentially producing an observable transient (light blue blob). Right: at
an age of around 100 kyr, crustal magnetic-field evolution above an
incomplete Meissner-expelled layer causes localised disruptions (red
field line) to an
otherwise charged-starved dipolar magnetosphere, resulting in radio
emission (orange cone) modulated by the star's rotation (green arrow).
}
\end{center}
\end{figure*}

The existence of an unmagnetised region in the core of NSs is crucial
to drive significant crustal field evolution within $10^4$ yr; without
any mechanism for flux expulsion, the currently accepted paradigm for
magnetar activity could be seriously undermined. Dissipative processes
are too slow to achieve the required Meissner effect
\cite{BPP69a,ho17}, but an advection-reconnection process could
produce an energetically favourable, flux-free, outer NS core on a
dynamical timescale \cite{meissner_long} and thus provide the first
theoretical support for this important assumption. Furthermore, this
mechanism could potentially be observed: through an energy injection
in the late supernova phase (see Fig. \ref{observables}), differences in NS rotational behaviour, and gravitational-wave emission from mature NSs.

Our simulations of NS crustal field evolution for the cases of partial or complete flux expulsion from the outer core imply that there are three
different observational incarnations of a highly magnetised NS. Only
one of these is a classical magnetar: the case of full Meissner
expulsion. Within the dynamical Meissner model this suggests their internal
fields are in the range $0.01\lesssim \bar{B}_{\rm{int},14}\lesssim
H_{c,14}\approx 5$. Interestingly, this is not very different from the
range of $B_{\rm dip}$ values inferred for known magnetars,
$0.06\lesssim B_\mathrm{dip,14}<8$ \cite{rea13}, with just one widely-quoted outlier value: $2\times 10^{15}\,\mathrm{G}$ for SGR 1806-20. This was, however, calculated during an epoch of enhanced spindown; the star has since stabilised to a lower spindown rate from which $B_{\rm dip}=7.7\times 10^{14}\,\mathrm{G}$ is estimated \cite{younes17}.

We predict that \skl{neither} NSs with
$\bar{B}_{\mathrm{int}}>H_c$, \skl{nor those with a substantial
  large-scale field after the birth phase, will} display magnetar activity,
although their substantial magnetic distortions could make them
promising sources of gravitational waves. Finally, NSs possessing a
partially Meissner-expelled shell with `holes' become active
once substantially older than typical magnetars, potentially powering some newly discovered radiative events that are otherwise theoretically challenging to explain. The growing number of very-long-period radio transients, and the expected discovery of many well-localized FRBs by upcoming instruments, will provide critical constraints on the existence of the `old magnetar' class we predict, and reveal details of their properties that can be used to better understand NS interiors.

\section{Methods}

\subsection{The onset of superconductivity}
The critical temperature $T_c$ for proton superconductivity varies throughout the core of a NS, peaking
at a value $\sim (3-7)\times 10^9$ K \cite{ho15}, which is attained
only minutes after the star's formation. The range of peak values for $T_c$ reflects differing approaches to calculating the energy
gap. Here we adopt an approximation \cite{ho12} to one particular
energy gap model \cite{chen93}, yielding $T_c$ as a function of mass
density $\rho$. We find a relativistic hydrostatic
equilibrium by solving the TOV equations with the SLy4 equation of state
\cite{DH01}, giving us $\rho$ and hence $T_c$ as a function of radius. Finally, combining this with a standard cooling
prescription \cite{page06} allows us to track the expansion of the
superconducting shell as a function of time. This gives a
first thin shell of superconductivity forming at a radius of $0.79R_*$, a temperature $T=6.8\times 10^9\,\mathrm{K}$, and a stellar age of $170$ seconds. The shell
spreads rapidly, enveloping half of the core radius (in the absence of
magnetic-field effects) after $\sim1$~day. Another important
quantity is $H_c$, which varies by a factor of $\sim 5$
within the core \cite{GAS}, but not monotonically, and also depends
upon the equation of state, the nature of core superconductivity, and
the geometry of $\bB$. We cannot account for all these features and
retain any generality in the model, so for simplicity we approximate
$H_c$ as a constant for our calculations, adopting a representative value $H_c=5\times
10^{14}$~G.

\subsection{Magnetic field evolution in NSs}

Magnetic field evolution in the NS core is complex, and the timescale of the dominant mechanisms remains contentious (see e.g. \cite{gusakov20} and references therein). This, in turn, makes it difficult to attribute any NS activity unambiguously to evolution in the core. By contrast, the crust is conceptually simple. Completely generally, the electric current is due to a relative flow of positive and negative charges. In the crust the positive charges are ions locked in a crystalline lattice, so that only the electrons are mobile. The electric current, and therefore $\bB$, depend only on the electron velocity, and from this restriction one may derive the equation
\begin{equation}\label{eMHD}
    \pd{\bB}{t}=-\frac{1}{4\pi}\curl\brac{
    \frac{c}{\rho_e}(\curl\bB)\times\bB
    -\frac{c^2}{\sigma}\curl\bB
    },
\end{equation}
where the first term on the right-hand side is Hall drift, the second
Ohmic decay, and $\rho_e$ is the charge density \cite{gold_reis92}.
Taking a simple poloidal field as the initial condition, we evolve equation \eqref{eMHD} numerically, specialising to axisymmetry for simplicity. We use a finite difference method, discretizing the numerical domain in $r$ and $\cos \theta$. The resolution is typically $100 \times 100$ but we have experimented in several cases with doubling the resolution, to confirm the numerical convergence of our results. 
We use a central-difference scheme for spatial derivatives, and three- and five-point stencils for second and third derivatives, respectively. The second order Adams-Bashford method is used for numerical time integration. The timestep is dynamically adapted using the Courant condition:
\begin{equation}
    \Delta t  \leq \frac{\Delta r}{v_{max}}
\end{equation}
where $\Delta r$ is the grid spacing and $v_{max}$ the maximum
electron velocity.
The profiles for $\rho_e$ and $\sigma$, and the numerical code used
here, are the same as in \cite{GL21}, except that for the present work
we do not allow for plastic flow.

The outcome of the dynamical Meissner process depends on the geometry
of the star's magnetic field and any fluid flows present at the onset of
superconductivity. For simplicity, however, we assume a
large-scale flow advecting an initially dipolar field. For this scenario, we use three families of
boundary condition at the crust-core interface: a fully expelled
magnetic field, where the radial field $B_r$ at the inner boundary is
zero; a magnetic field containing two holes of semi-opening angle
$\arccos{0.8}\approx 37^{\circ}$ where the radial field is non-zero;
and a boundary condition where there is a non-zero radial field in a
band at the equator, starting at $\arccos{0.3}=73^{\circ}$ and ending
at $\arccos{-0.3}=108^{\circ}$. In all three cases the radial field at
the inner boundary is static, but the tangential field components
($B_\theta$ and $B_\phi$) are free to evolve. At the outer surface of
the crust and into the exterior, we assume a vacuum magnetic field,
i.e. no electric current, using a multipole fit which serves as the
external boundary condition. 
Note that although our boundary
conditions are based on simplified axisymmetric endpoints predicted by
the dynamical Meissner effect \cite{meissner_long}, any other
mechanism resulting in partial flux expulsion might induce similar
long-term field evolution.

The three evolutions, whose endpoints at 0.1 Myr are plotted in
  Fig. \ref{crust_B}, proceed as follows. All three show significant
  evolution and the development of a toroidal-field component in the
  outer crustal region within 1 kyr; this component begins to extend
  deeper into the crust up until an age of $\sim 10$ kyr. The two partially-expelled cases
  also show a corresponding development of toroidal field in the
  innermost part of the crust, at the boundaries of the holes where
  field lines pass into the core. In these two cases, after $\sim 10$ kyr the inner and outer
  regions with toroidal field meet in the middle and interact, and the
  toroidal component becomes more intense, reaching a peak value just above
  the holes and at an age of around 50 kyr. There is also considerable
  disruption to the poloidal-field component above this region, with
  small loops of magnetic field rising up out of the crust. This
  activity is sustained right up until the end of our simulations, at
  0.1 Myr, so would clearly continue beyond this point, although
  numerical restrictions prevent us from following it further. By
  contrast, the standard boundary condition, of field expelled from
the core, shows limited evolution beyond $\sim 20$ kyr, with the field
having relaxed to a quasi-equilibrium state by the end of the
simulation, at 0.1 Myr.

\subsection{Data and code availability}

Both the code used to evolve equation \eqref{eMHD}, and the data
generated from these evolutions -- as presented in Fig.
\ref{crust_B} -- are available upon reasonable request.
Correspondence and requests for materials should be addressed to the
first author.

\begin{acknowledgments}
This work was supported by computational time granted from the
National Infrastructures for Research and Technology S.A. (GRNET S.A.)
in the National HPC facility-ARIS-under project ID
pr017008/simnstar2. We thank Bryn Haskell, Mike J. Moss and Brad Cenko
for helpful discussions on some of these topics. Z.W. acknowledges support by NASA under award number 80GSFC21M0002 and 80GSFC24M0006. D.A. acknowledges support from a UKRI (STFC/EPSRC) fellowship (EP/T017325/1). This work made use of the NASA Astrophysics Data System. 
\end{acknowledgments}

SKL formulated the model, explored its basic consequences, and took the
primary role of writing much of the paper; KNG implemented the new
boundary conditions in a code and performed the numerical simulations; ZW and DA
explored some of the observational consequences in detail; DA also
edited and proofread the paper.


\bibliographystyle{apsrev4-2}
\bibliography{references}


\label{lastpage}

\end{document}